\pdfoutput=1
\documentclass[12pt]{article}
\usepackage{amsmath}
\usepackage{mathrsfs}
\usepackage{amsfonts}
\usepackage{natbib}
\usepackage{multirow}
\usepackage{enumitem}
\usepackage{amsthm}
\usepackage{amssymb}
\usepackage{color}
\usepackage{lscape}
\usepackage{rotating}
\usepackage{caption}
\usepackage{subcaption}
\usepackage{graphicx}
\usepackage{xcolor}
 
\newcommand{\Date}[1]{\def\@Date{#1}}
\def\today{\number\day~\ifcase\month\or
 January\or February\or March\or April\or May\or June\or
 July\or August\or September\or October\or November\or December\fi~\number\year}

\def\be{\begin{equation}}
\def\ee{\end{equation}}
\def\bea{\begin{eqnarray}}
\def\eea{\end{eqnarray}}
\def\bd{\begin{displaymath}}
\def\ed{\end{displaymath}}
\def\bda{\begin{eqnarray*}}
\def\eda{\end{eqnarray*}}
\def\bsm{\begin{small}}
\def\esm{\end{small}}

\def\t0{\theta_0}
\def\nn{\nonumber}

\def\ha1{\widehat \beta_1}

\def\bnt{\begin{enumerate}}
\def\ent{\end{enumerate}}
\def\T{{ \mathrm{\scriptscriptstyle T} }}

\def\oc{{ \mathrm{\scriptstyle or} }}
\def\ps{{ \mathrm{\scriptstyle ps} }}

\def\bsc{\begin{scriptsize}}
\def\esc{\end{scriptsize}}

\newtheorem{theorem}{Theorem}

\newtheorem{proposition}{Proposition}
\newtheorem{condition}{Condition}

\theoremstyle{definition}

\newcommand{\E}{\mathbb{E}}
\newcommand{\PP}{\mathbb{P}}
\newcommand{\V}{{\rm Var}}

\makeatletter
\newcommand{\figcaption}{\def\@captype{figure}\caption}
\newcommand{\tabcaption}{\def\@captype{table}\caption}
\makeatother

\newcommand{\argmin}{{\rm argmin}}

\newcommand{\cov}{{\rm Cov}}

\newcommand{\bA}{{\mathbf A}}

\newcommand{\bM}{{\mathbf M}}

\newcommand{\bU}{{\mathbf U}}

\newcommand{\bX}{{\mathbf X}}

\newcommand{\bZ}{{\mathbf Z}}
\newcommand{\ba}{{\mathbf a}}
\newcommand{\bb}{{\mathbf b}}

\newcommand{\bv}{{\mathbf v}}

\newcommand{\bx}{{\mathbf x}}

\newcommand{\bbeta}  {\boldsymbol{\beta}}

\newcommand{\blambda}{\boldsymbol{\lambda}}
\newcommand{\bepsilonb}{\boldsymbol{\varepsilon}}

\newcommand{\bSigma}{\boldsymbol{\Sigma}}
\newcommand{\bDelta}{\boldsymbol{\Delta}}
\newcommand{\bgamma}{\boldsymbol{\gamma}}

\newcommand{\bpi}{\boldsymbol{\pi}}

\newcommand{\bmu} {\boldsymbol{\mu}}

\newcommand{\bzero}{{\mathbf 0}}

\usepackage{accents}

\newcommand{\blind}{1}

\begin{document}
\def\spacingset#1{\renewcommand{\baselinestretch}%
{#1}\small\normalsize} 
\spacingset{1}

\if1\blind
{
  \title{\bf 
  Multiply Robust Inference of Average Treatment Effects by High-dimensional Empirical Likelihood}
  \author{Xintao Xia \\
  Center for Data Science, Zhejiang University, Zhejiang, China\\
  and \\
  Yumou Qiu \\
  School of Mathematical Sciences, Peking University, Beijing, China  
  }
\date{}
\maketitle
} \fi

\bigskip

\begin{abstract}
In this paper, we develop a multiply robust inference procedure of the average treatment effect (ATE) for data with high-dimensional covariates. We consider the case where it is difficult to correctly specify a single parametric model for the propensity scores (PS). For example, the target population is formed from heterogeneous sources with different treatment assignment mechanisms. We propose a novel high-dimensional empirical likelihood weighting method under soft covariate balancing constraints to combine multiple working PS models. An extended set of calibration functions is used, and a regularized augmented outcome regression is developed to correct the bias due to non-exact covariate balancing. Those two approaches provide a new way to construct the Neyman orthogonal score of the ATE. The proposed confidence interval for the ATE achieves asymptotically valid nominal coverage under high-dimensional covariates if any of the PS models, their linear combination, or the outcome regression model is correctly specified. The proposed method is extended to generalized linear models for the outcome variable. Specifically, we consider estimating the ATE for data with unknown clusters, where multiple working PS models can be fitted based on the estimated clusters. Our proposed approach enables robust inference of the ATE for clustered data.
We demonstrate the advantages of the proposed approach over the existing doubly robust inference methods under high-dimensional covariates via simulation studies.
We analyzed the right heart catheterization dataset, initially collected from five medical centers and two different phases of studies, to demonstrate the effectiveness of the proposed method in practice. 

\end{abstract}

\noindent%
\vfill

\newpage
\spacingset{1.9}

\section{Introduction}
\label{sec:intro}

The estimation of the average treatment effect stands as one of the fundamental problems in causal inference. The concept of propensity score \citep{rosenbaum1983central} is central in addressing and mitigating potential selection biases in non-randomized trials. Estimated propensity scores are used in various methods for estimating ATEs, including PS matching, the inverse probability weighted (IPW) estimator, also known as the Horvitz–Thompson estimator, and the augmented IPW (AIPW) estimator that incorporates the outcome regression (OR) model, thereby attaining the doubly robust and semi-parametrically efficient properties \citep{bang2005doubly}. Covariate balancing, which uses the IPW formulation with the estimated PS under the covariate balancing constraints, is another way to estimate the ATEs \citep{imai2014covariate}. The AIPW and covariate balancing estimators are asymptotically equivalent under the fixed-dimensional setting, and their properties have been well studied in this regime. See, for example, \cite{cao2009improving, imai2014covariate}. 

The well-known doubly robust procedures only utilize a single PS and OR model. The two models are likely misspecified when the data are high-dimensional and from multiple heterogeneous sources with different treatment assignment mechanisms. For example, estimating ATEs based on combined datasets is common in clinical trials and medical studies. The right heart catheterization (RHC) dataset \citep{connors1996effectiveness} was collected from five medical centers and two studies, a prospective observational study and an interventional study, to examine the effect of the RHC treatment on the survival rate of severely ill patients. However, the released data do not share the subgroup information. Given the unknown subgroup structure, a single PS model is unlikely to be adequate. 

\subsection{Related work}

In fixed-dimensional settings, \cite{Han2014} proposed a multiply robust inference method by empirical likelihood (EL) that allows multiple working models for both the propensity score and outcome regression. Valid statistical inference can be obtained if any of those models are correctly specified. \cite{chan2014oracle} explored generalized empirical likelihood calibration estimators to achieve multiple robustness for missing data. \cite{chen2017multiply} proposed a multiply robust imputation method in survey sampling settings. However, those methods do not consider high-dimensional covariates or data with clusters. 

In high-dimensional settings, \cite{belloni2014inference} and \cite{farrell2015robust} introduced a double selection approach for estimating the ATE. Their approaches employ regularized estimators for both the PS and OR models in the AIPW formulation. \cite{chernozhukov2018double} developed a versatile and flexible framework for conducting inference on a target parameter under high-dimensional nuisance parameters. Their method utilizes the Neyman orthogonal score and data-splitting technique, which includes the AIPW estimator as a special case for estimating the ATE. These methods are doubly robust for estimation of the ATE. However, these methods do not yield doubly robust inference, as valid inference requires correct specification of both the PS and OR models. In a related effort to derive asymptotically valid confidence intervals for ATEs, \cite{athey2018approximate} required correct specification of a linear outcome model. In recent work on linear outcome models, \cite{ning2020robust, tan2020model} proposed doubly robust inference methods for ATEs under high-dimensional data by regularized calibrated estimation. Their proposed confidence intervals of ATEs achieve an asymptotically valid coverage if either the PS or OR model is correctly specified. Particularly, \cite{ning2020robust} also considered generalized linear models (GLM) for the outcome variable. 

All the aforementioned high-dimensional methods do not utilize multiple PS models. The performance of doubly robust estimators can deteriorate, even with slight misspecification in both the PS and OR models. The phenomenon was highlighted in a comprehensive set of numerical studies by \cite{kang2007demystifying}, which demonstrated that, in certain scenarios, using two incorrectly specified models does not yield better results than using a single misspecified model. Therefore, it is important to develop multiply robust inference methods under high-dimensional data, especially when we lack confidence in using a single propensity score model to capture complex treatment assignment mechanisms.

\subsection{Our contributions}

In this paper, we develop a novel method for estimating ATEs that utilizes multiple PS models with high-dimensional covariates. The proposed method achieves multiply robust inference under high-dimensional data; that is, it provides asymptotically valid statistical inference for ATEs if any working PS model, their linear combination, or the outcome regression model is correctly specified. Moreover, the proposed estimator achieves semiparametric efficiency if both the OR model and one of the working PS models (or their linear combination) are correctly specified. The multiply robust inference property is important for clustered data from heterogeneous sources, as a single PS model may not be adequate to reflect the true treatment assignment mechanism. Combining multiple working PS models based on the estimated clusters provides more guarantees of achieving valid statistical inference of the ATE against misspecified clustering algorithms or PS models under a heterogeneous population with subgroup structures. 

Our approach comprises high-dimensional empirical likelihood weighting, regularized augmented outcome regression, and high-dimensional clustering. We develop a novel soft covariate balancing procedure by high-dimensional EL to combine estimated propensity scores from multiple working PS models. Our soft calibration EL is related to the regularized EL estimation in \cite{chang2018new, chang2021high}. However, the main goals are different. The regularized EL approach is constructed to estimate high-dimensional estimating equations, while our method is proposed to estimate the propensity scores that achieve approximate covariate balancing. Note that exact covariate balancing is not possible under high-dimensional covariates. To achieve the multiply robust inference property, we apply soft calibration to derivatives of all working PS models and fit augmented outcome regression with regularization to correct the bias due to non-exact covariate balancing. These two approaches provide a new way to construct the Neyman orthogonal score of the ATE, ensuring that the estimation variability of all nuisance parameters does not affect the influence function of the estimated ATE. In this sense, our proposed method is also related to the existing literature on high-dimensional inference of regression models \citep{zhang2014confidence, van2014asymptotically}. The proposed procedure is also extended to GLMs for the outcome variable. Simulation studies demonstrate the superiority of the proposed method over the existing methods. 

\subsection{Organization and notations}

This paper is organized as follows. Section \ref{sec:pre} introduces the problem formulation of estimating ATEs under high-dimensional data and the setting of multiple robust inferences. Section \ref{sec:lm_method} presents the proposed multiply robust procedure under high-dimensional covariates. Section \ref{sec:lm_thm} delves into the theoretical properties. Section \ref{sec:glm} extends the proposed approach to encompass GLMs for the outcome regression. Section \ref{sec:group} considers the estimation of ATEs under clustered data. Section \ref{sec:num} conducts simulation studies to verify the proposed method and compare it with the existing doubly robust procedures under high dimensional data. Section \ref{sec:realdata} analyzes the treatment effect of RHC using a real data set with unknown clusters. Section \ref{sec:con} summarizes the paper. 
A simplified proof to explain the multiply robust property of the proposed method is presented in the Appendix, which explains the importance of soft calibration using augmented covariates and augmented outcome regression to achieve multiply robust inferences under high-dimensional covariates. 
The theoretical results under GLM, additional numerical results, and all technical proofs are relegated to the supplementary material (SM).

For a $p$-dimensional vector $\bx=(x_1,\dots,x_p)^{\T}$, let $\|\bx\|_q = (\sum_{j = 1}^{p}|x|_j^q)^{1/q}$ and $\|\bx\|_{0} = \sum_{j = 1}^{p} 1(x_j \neq 0)$ be the $\ell_q$ and $\ell_0$ norms of $\bx$, respectively, where $1(\cdot)$ denotes the indicator function. For a $p\times p$ symmetric positive semidefinite matrix $\bA$, let $\lambda_{\text{max}}(\bA)$ and $\lambda_{\text{min}}(\bA)$ be the maximal and minimal eigenvalues of $\bA$, respectively. For two sequences of random variables, $\{a_n\}$ and $\{b_n\}$, $a_n \approx b_n$ and $a_n=o_p(b_n)$ denote $|a_n-b_n|$ and $a_n/b_n$ converging to zero in probability, respectively. For two sequences of positive real numbers, $\{\tilde{a}_n\}$ and $\{\tilde{b}_n\}$, we use $\tilde{a}_n \asymp \tilde{b}_n$ to indicate that $\tilde{c}_1 \leq a_n / b_n \leq \tilde{c}_2$ for two positive constants $\tilde{c}_1$ and $\tilde{c}_2$.

\setcounter{equation}{0}
\section{Problem and Settings}
\label{sec:pre}

Let $\bX = (X_{1}, \ldots, X_{p})^{\T}$ be the $p$-dimensional covariate and $D$ be the indicator of treatment assignment, taking the value $1$ if the observation receives the treatment and $0$ otherwise. Let $Y(d)$ be the potential outcome under $D = d$ for $d = 0, 1$, and $Y = D Y(1) + (1 - D) Y(0)$ be the observed response variable. Let $\mathcal{G}$ be the joint distribution of $(D, \bX^{\T}, Y(0), Y(1))$. Suppose the observed samples $\{(D_i, \bX_i^{\T}, Y_i): i=1,\ldots,n\}$ are independent and identically distributed (IID) realizations from the joint distribution of $(D, \bX^{\T}, Y)$. Let $n_1 = \sum_{i = 1}^{n} D_i$ and $n_0 = n - n_1$ be the sample sizes of the treated and control groups, respectively. We consider the case of high-dimensional covariates where $p$ can be much larger than $n$. Let $\mu_1 := {\E} \{ Y(1) \}$ and $\mu_0 := {\E} \{ Y(0) \}$, where the expectation is taken with respect to the population distribution $\mathcal{G}$. We are interested in estimating the \textit{average treatment effect}:
\begin{equation} \label{eq:est_eq}
\tau := \mu_1 - \mu_0 = {\E} \{ Y(1) - Y(0) \}.
\end{equation}

To identify both $\mu_0$ and $\mu_1$, we assume that the potential outcomes and treatment assignment are conditionally independent given the covariates, as formalized in the following unconfoundedness assumption.

\begin{condition}[Unconfoundedness]\label{as:ignorability}
    $Y(1)$ and $Y(0)$ are conditionally independent of $D$ given $\bX$ such that $Y(d) \perp D \mid \bX$ for $d = 0, 1$.
\end{condition}

Under Condition \ref{as:ignorability}, the parameter $\mu_d$ can be identified by either the outcome regression of $\E(Y \mid \bX, D = d)$ or inverse weighting of the propensity score $\pi(\bX) := \PP(D = 1 \mid \bX)$. The doubly robust inference methods \citep{tan2020model, ning2020robust} provide valid statistical inference of ATE when either the OR or PS model is correctly specified. However, specifying a single correct PS model may be challenging, especially when the target population contains multiple heterogeneous subgroups. We consider the problem of multiply robust inference \citep{Han2014, Li2020Demystifying} in high-dimensional settings, where the true PS model is assumed to lie within the convex hull of multiple candidate models $\{\pi_k(\bX, \bgamma_k): k = 1, \ldots, q\}$ such that 
\be\label{eq:MPS}
\pi(\bX) = \sum_{k = 1}^{q} d_k \pi_k(\bX, \bgamma_k) \ee
for unknown parameters $\{\bgamma_k\}_{k=1}^{q}$ and weights $\{d_k\}_{k=1}^{q}$ where $d_k \geq 0$. A mixture model of conditional density, similar to the setting in (\ref{eq:MPS}), is considered for distributionally robust prediction \citep{wang2023distributionally}. Let $\bb_{\oc}(\bX)=(b_0(\bX), \ldots, b_{p_0}(\bX))^{\T}$ be a set of $p_0$ functions of $\bX$, where $p_0$ could be larger than $p$ and much larger than $n$. We consider a working linear OR model for ${\E}\{Y(d) \mid \bX\}$ such that
\be\label{eq:OR-model}
{\E}\{Y(d) \mid \bX\} \subset \mbox{span}\{b_0(\bX), \ldots, b_{p_0}(\bX)\}, \mbox{ \ where \ } b_0(\bX) = 1,
\ee
where $\mbox{span}\{b_0(\bX), \ldots, b_{p_0}(\bX)\}$ denotes the linear space spanned by $\bb_{\oc}(\bX)$. A multiply robust procedure provides asymptotically valid statistical inference of the ATE if either the OR model in (\ref{eq:OR-model}) is correctly specified or any one of the PS models $\{\pi_k(\bX, \bgamma_k)\}_{k = 1}^{q}$ or their linear combination in (\ref{eq:MPS}) is correctly specified. 

The multiply robust procedure is more advantageous than the doubly robust procedure, which only use a single PS model. In many scientific studies, datasets often combine studies from multiple sources. See the right heart catheterization dataset \citep{connors1996effectiveness} analyzed in Section \ref{sec:realdata}. The multiply robust property is beneficial particularly when the target population comes from multiple heterogeneous sources and the origin of each observation is unknown. Section \ref{sec:group} introduces the ATE estimation under clustered data, where combining multiple propensity score estimates based on clustering results enables valid statistical inference. The existing high-dimensional doubly robust inference methods cannot be applied to multiple PS models, and the existing multiply robust inference methods are designed only for fixed-dimensional settings.  

\setcounter{equation}{0}
\section{Methodology}
\label{sec:lm_method}

We present the proposed multiply robust method to estimate $\mu_1={\E}\{Y(1)\}$ under high-dimensional covariates. The estimator of $\mu_0={\E}\{Y(0)\}$ can be similarly constructed. Let $\pi_{k}^{(1)}(\bX, \bgamma_k) = \partial \pi_{k}(\bX, \bgamma_k) / \partial \bgamma$ be the first-order derivative of the working PS model $\pi_k(\bX, \bgamma_k)$, and $\widehat{\bgamma}_k$ be an estimate of $\bgamma_k$ for $k = 1, \dots, q$. Let 
\be
\widehat{\bb}(\bX) = (b_1(\bX), \ldots, b_{p_0}(\bX), \pi_{1}^{(1)}(\bX, \widehat{\bgamma}_1), \ldots, \pi_{q}^{(1)}(\bX, \widehat{\bgamma}_q))^{\T}
\label{eq:basis}\ee 
be the augmented covariate functions, where we use $\widehat{b}_j(\bX)$ to denote $j$th component of $\widehat{\bb}(\bX)$. The accent in $\widehat{\bb}(\bX)$ indicates the dependence on the estimates $\{\widehat{\boldsymbol{\gamma}}_k\}_{k=1}^q$. To keep the notations consistent, we also use $\widehat{b}_j(\bX)$ to denote the first $p_0$ components of $\widehat{\bb}(\bX)$, even though they are free of estimates. Let $\overline{\bb} = (\bar{b}_1, \ldots, \bar{b}_m)^{\T} = n^{-1} \sum_{i = 1}^{n} \widehat{\bb}(\bX_i)$ be the overall sample mean and $\widetilde{\bb}_i = (\widetilde{b}_{i1}, \ldots, \widetilde{b}_{im})^{\T} = \widehat{\bb}(\bX_i) - \bar{\bb}$. The treatment group empirical likelihood solves the constrained optimization problem:
\bea
\{\widetilde{p}_i: D_i = 1\} &=& \underset{p_i}{\operatorname{argmax}} \sum_{i = 1}^{n} D_i \log(p_i), \mbox{ \ subject to \ } p_i \geq 0, \ \sum_{i = 1}^{n} D_i p_i = 1, \ \ \ \ \ \label{eq:EL-observe} \\
&& \sum_{i = 1}^{n} D_i p_i \{\pi_k(\bX_i, \widehat{\bgamma}_k) - n_1 / n\} = 0 \mbox{ \ for $k = 1, \ldots, q$ and } \label{eq:EL-observe-IBC} \\
&& \bigg| \sum_{i = 1}^{n} p_i D_i \widetilde{b}_{ij} \bigg| \leq \omega_{\ps} \mbox{ \ for $j = 1, \ldots, m$}, \label{eq:EL-observe-calibration}
\eea
where (\ref{eq:EL-observe-IBC}) and (\ref{eq:EL-observe-calibration}) are called internal bias correction (IBC) constraints and soft calibration (soft covariate balancing) constraints, respectively. Here, soft calibration refers to the property that the weighted mean $\sum_{i} D_i \tilde{p}_i \widehat{b}_{j}(\bX_i)$ of $\{\widehat{b}_{j}(\bX_i)\}_{i=1}^{n}$ over the treatment group is approximately, though not necessarily exactly, equal to the overall sample mean $\bar{b}_j$ for $j=1,\dots,m$, and $\omega_{\ps}$ is the tolerance parameter for non-exact covariate balancing. Calibrating the derivatives of the PS models is important for the multiply robust inference of the ATE. See the explanations from (\ref{eq:expansion}) to (\ref{eq:outcome-bound-2}) for the asymptotic expansion of the proposed estimator under the misspecified outcome regression model. In the fixed-dimensional setting, \cite{Han2014} proposed EL weighting with exact covariate balancing. However, in the high-dimensional setting, exact balancing is generally infeasible, as no solution exists for the weights $\{\tilde{p}_i\}_{i=1}^{n}$ that simultaneously satisfy all exact balancing constraints.

Let $\widehat{\bpi}_i = (\pi_1(\bX_i, \widehat{\bgamma}_1), \ldots, \pi_q(\bX_i, \widehat{\bgamma}_q))^{\T}$ be the estimated propensity scores for individual $i$ from $q$ working PS models, and $\mathbf{1}_q = (1, \ldots, 1)^{\T}$ be a $q$-dimensional vector of ones. The solutions of EL $\{\widetilde{p}_i: D_i = 1\}$ for the treatment group take the form
\be
n \widetilde{p}_i = \{\widetilde{\lambda}_0 + \widetilde{\blambda}_{1}^{\T} (\widehat{\bpi}_i - n_1 / n \mathbf{1}_q) + \widetilde{\blambda}_{2}^{\T} \widetilde{\bb}_i\}^{-1},
\label{eq:EL-weights}
\ee
where $\widetilde{\lambda}_{0}$, $\widetilde{\blambda}_{1} = (\widetilde{\lambda}_{11}, \ldots, \widetilde{\lambda}_{1q})^{\T}$, $\widetilde{\blambda}_{2} = (\widetilde{\lambda}_{21}, \ldots, \widetilde{\lambda}_{2m})^{\T}$ are Lagrange multipliers such that the solutions $\{\widetilde{p}_i: D_i = 1\}$ satisfy constraints in (\ref{eq:EL-observe}), (\ref{eq:EL-observe-IBC}) and (\ref{eq:EL-observe-calibration}). Using dual optimization, $(\widetilde{\lambda}_{0},\widetilde{\blambda}_{1},\widetilde{\blambda}_{2})$ can be obtained by the regularized unconstrained optimization problem:
\be
\underset{\lambda_{0}, \blambda_{1}, \blambda_{2}}{\operatorname{argmin}} \frac{1}{n} \sum_{i = 1}^{n} -D_i \log\{\lambda_0 + \blambda_{1}^{\T} (\widehat{\bpi}_i - n_1 / n \mathbf{1}_q) + \blambda_{2}^{\T} \widetilde{\bb}_i\} +\lambda_0+ \omega_{\ps} \|\blambda_{2}\|_1.
\label{eq:EL-pi}
\ee
By the Karush–Kuhn–Tucker (KKT) conditions for problem (\ref{eq:EL-pi}), it can be shown that the EL weights $\{\widetilde{p}_i\}$, obtained by plugging in the minimizers of (\ref{eq:EL-pi}), satisfy the constraints in (\ref{eq:EL-observe})--(\ref{eq:EL-observe-calibration}). The empirical likelihood weights $\{\widetilde{p}_i: D_i = 0\}$ of the control group can be obtained in a similar manner.

Under the fixed-dimensional setting with exact covariate balancing, namely $\omega_{\ps} = 0$ in (\ref{eq:EL-observe-calibration}), the weighted mean $\sum_{i = 1}^{n} D_i \widetilde{p}_i Y_i$ is an implicit AIPW estimator of $\mu_1$ \citep{Han2014}. However, the equivalence no longer holds in the high-dimensional setting due to the soft calibration in (\ref{eq:EL-observe-calibration}). To correct the bias from the non-exact covariate balancing, we propose to estimate an augmented outcome regression (AOR) of $Y$ under the treatment group. Let $\{\widetilde{\beta}_0,\widetilde{\bbeta}_1, \widetilde{\bbeta}_2\}$ be the minimizer of the regularized weighted linear regression:
\be
\underset{\beta_0,\bbeta_1, \bbeta_2}{\operatorname{argmin}} \frac{1}{n} \sum_{i = 1}^{n} \frac{D_i}{ \widetilde{\pi}_i^{2}} \{Y_i - \beta_0 - \bbeta_1^{\T} \widehat{\bb}(\bX_i) - \bbeta_2^{\T} \widehat{\bpi}_i \}^2 + \omega_{\oc} (\|\bbeta_1\|_1 + \|\bbeta_2\|_1) \ \ \
\label{eq:EL-outcome}
\ee
with the weights $\{\widetilde{\pi}_i^{-2}\}_{i=1}^{n}$ on the treatment group, where $\widetilde{\pi}_i = (n \widetilde{p}_i)^{-1}$ and $\omega_{\oc}$ is a penalty parameter. In addition to the covariates $\bb_{\oc}(\bX)$ under the working OR model in (\ref{eq:OR-model}), the augmented outcome regression also includes the working PS model $\pi_k(\bX_i, \widehat{\bgamma}_k)$ and its derivatives $\pi_{k}^{(1)}(\bX, \widehat{\bgamma}_k)$ as additional covariates. The regression coefficient $\bbeta_1$ is composed as $\bbeta_1 = (\bbeta_{10}^{\T}, \bbeta_{11}^{\T}, \ldots, \bbeta_{1q}^{\T})^{\T}$, where $\bbeta_{10}$ and $\bbeta_{1k}$ are the coefficients of covariates $\{b_j(\bX)\}_{j = 1}^{p_0}$ and $\pi_{k}^{(1)}(\bX, \widehat{\bgamma}_k)$ for $k=1,\dots,q$, respectively. The proposed multiply robust estimator of $\mu_1$ using the AIPW formulation is
\be
\widehat{\mu}_1 = \frac{1}{n}\sum_{i = 1}^{n} \frac{D_i Y_i}{\widetilde{\pi}_i} - \frac{1}{n} \sum_{i = 1}^{n} \frac{D_i - \widetilde{\pi}_i}{\widetilde{\pi}_i} \{\widetilde{\beta}_0+\widetilde{\bbeta}_1^{\T} \widehat{\bb}(\bX_i) + \widetilde{\bbeta}_2^{\T} \widehat{\bpi}_i\}.
\label{eq:Est-mu-1}
\ee
The multiply robust estimator $\widehat{\mu}_0$ of $\mu_0$ can be similarly constructed. The proposed estimate of ATE is $\widehat{\tau} = \widehat{\mu}_1 - \widehat{\mu}_0$.

The existing approaches for high-dimensional causal inference \citep{belloni2014inference, ning2020robust, tan2020model} performed calibration only for the covariates of the PS model and apply regularized regression only to the covariates of the OR model. However, such strategies are not sufficient to guarantee the multiple robustness of the estimator. Incorporating augmented covariates in both soft calibration and outcome regression induces Neyman orthogonality for all nuisance parameters, so that the statistical inference of the ATE is multiply robust against misspecified PS models $\{\pi_k(\bX, \bgamma_k)\}_{k = 1}^{q}$ in (\ref{eq:MPS}) or misspecified OR model in (\ref{eq:OR-model}). We explain the importance of calibrating derivatives of propensity scores and augmented outcome regression in the Appendix. 

\setcounter{equation}{0}
\section{Theoretical Results}
\label{sec:lm_thm}

In this section, we provide conditions to show the multiply robust properties of the proposed estimator. Let $\{\bgamma_{k, \ast}\}_{k = 1}^{q}$, $\lambda_{0, \ast}$, $\blambda_{1, \ast}$, $\blambda_{2, \ast}$, $\beta_{0, \ast}$, $\bbeta_{1, \ast}$, $\bbeta_{2, \ast}$ be the probability limits of $\{\widehat{\bgamma}_k\}_{k = 1}^{q}$, $\widetilde{\lambda}_{0}$, $\widetilde{\blambda}_{1}$, $\widetilde{\blambda}_{2}$, $\widetilde{\beta}_0$, $\widetilde{\bbeta}_{1}$, $\widetilde{\bbeta}_{2}$, as $n, p \to \infty$, respectively, under either correctly specified or misspecified models. The rigorous definitions of these parameters are given in Section S3 of the SM. Let $\bpi_{i, \ast} = (\pi_1(\bX_i, \bgamma_{1, \ast}), \ldots, \pi_q(\bX_i, \bgamma_{q, \ast}))^{\T}$, $\bb_{\ast}(\bX_i) = (b_1(\bX), \ldots, b_{p_0}(\bX), \pi_1^{(1) }(\bX_i,\bgamma_{1,*})^{\T},\dots,\pi_q^{(1)}(\bX_i,\bgamma_{q,*})^{\T})^{\T}$ and $\blambda_{\ast} = (\lambda_{0, \ast}, \blambda_{1, \ast}^{\T}, \blambda_{2, \ast}^{\T})^{\T}$. Let $s_1=\max_{1\leq k\leq q}\|\bgamma_{k,*}\|_0$ and $s_2=\|(\bbeta_{1,*}^{\T},\bbeta_{2,*}^{\T})\|_0$ denote the sparsity levels of the propensity score models and the outcome regression model, respectively. Let $\pi_{k, j}^{(1)}(\bX_i,\bgamma_k)$ denote the $j$th component of $\pi_{k}^{(1)}(\bX_i,\bgamma_k)$. We make the following regularity conditions.

\begin{condition}\label{as:overlap}
   There exists a positive constant $c_0\in(0,1/2)$ such that the true propensity score $\pi_i = \pi(\bX_i)$ satisfies $c_0 \leq \pi_i \leq 1-c_0$ for $1\leq i \leq n$.
\end{condition}

\begin{condition}\label{as:sub-gaussian-design}
   The random variables $\bb_{\ast}(\bX)$ and $\epsilon_{1} = Y(1) - \beta_{0,*} - \bbeta_{1,*}^{\T}\bb_{\ast}(\bX)-\bbeta_{2,*}^{\T}\bpi_{*}$ are sub-Gaussian distributed. Namely, there is a positive constant $C$, such that for every $t \geq 0$ and $\bv\in\mathbb{R}^{m}$ satisfying $\|\boldsymbol{v}\|_2=1$, we have $\PP(|\bv^{\T} \bb_{\ast}(\bX)| \geq t)\leq 2\exp(-t^2/C^2)$ and $\PP(|\epsilon_1|\geq t)\leq 2\exp(-t^2/C^2)$. When the outcome model is misspecified, the random variable $\epsilon_{1}$ is bounded.
\end{condition}

\begin{condition}\label{as:eigen}
   The minimum and maximum eigenvalues of $\bSigma = \cov\{(\bb_{\ast}^{\T}(\bX),\bpi_{*})^{\T}\}$ satisfy $C\leq\lambda_{\text{min}}(\bSigma)\leq\lambda_{\text{max}}(\bSigma)\leq 1/C$ for a positive constant $C>0$.
\end{condition}

\begin{condition}\label{as:est_prop}
    (i) The propensity score models satisfy $\mathbb{E}\{\pi_k(\bX_i,\bgamma_{k,*})\} = \PP(D = 1)$, $c_0\leq\pi_k(\bX_i,\bgamma_{k,*})\leq 1-c_0$ for a positive constant $c_0\in(0,1/2)$, and are Lipschitz continuous $|\pi_k(\bX_i,\bgamma_{k})-\pi_k(\bX_i,\bgamma_{k}^{\prime})|\leq C|\bX_i^{\top}(\bgamma_{k,*}-\bgamma_{k}^{\prime})|$, $|\pi_{k,j}^{(1)}(\bX_i,\bgamma_{k})-\pi_{k,j}^{(1)}(\bX_i,\bgamma_{k}^{\prime})|\leq C|\bX_i^{\top}(\bgamma_{k,*}-\bgamma_{k}^{\prime})|$ for all $j$ and $|\bX_i^{\top}(\bgamma_{k,*}-\bgamma_{k}^{\prime})|=o_p(1)$. (ii) Estimates $\{\widehat{\bgamma}_k\}_{k=1}^{q}$ satisfy $\max_{k}\|\widehat{\bgamma}_k - \bgamma_{k,\ast}\|_1\leq Cs_1\{\log (p)/n\}^{1/2}$, $\max_{k}\|\widehat{\bgamma}_k - \bgamma_{k,\ast}\|_2\leq C\{s_1(\log p)/n\}^{1/2}$, $|(\bgamma_{k,*}-\widehat{\bgamma}_k)^{\top}\pi_{k}^{(2)}(\bX_i,\bDelta)(\bgamma_{k,*}-\widehat{\bgamma}_k)|\leq C|\bX_i^{\top}(\bgamma_{k,*}-\widehat{\bgamma}_k)|^2$ for $\bDelta\in[\bgamma_{k,*},\widehat{\bgamma}_k]$ with probability approaching to 1.
    (iii) If all the propensity score models are misspecified and (\ref{eq:MPS}) does not hold, then the population value $\blambda_{\ast}$ of the optimization problem \eqref{eq:EL-pi}, defined in (S3.8) in the SM, satisfies $\|\blambda_{2,*}\|_0=s_{\lambda}$, $s_1^{1/2}s_{\lambda}\log(m)=o(n^{1/2})$, and  $c_0\leq\pi_i^*\leq 1-c_0$ for a positive constant $c_0\in(0,1/2)$.
\end{condition}

\begin{condition}\label{as:sparsity}
   Let $s_0=\max\{s_1,s_2\}$. Sparsity parameters satisfy $(s_0s_1s_2)^{1/2} \log(m)=o(n^{1/2})$.
\end{condition}

Condition \ref{as:overlap} is the overlapping condition of the treatment and control groups, which is commonly required for causal inference. Conditions \ref{as:sub-gaussian-design} and \ref{as:eigen} are conventional assumptions in high-dimensional analysis \citep{van2014asymptotically, farrell2015robust}, which are needed for concentration inequalities and the compatibility condition for high-dimensional regularized regression. The assumption of bounded errors $\epsilon_{1}$ under a misspecified outcome model can be relaxed to a sub-Gaussian error if a stronger sparsity condition, $(s_0s_1s_2)^{1/2} \log(m)\log(n)=o(n^{1/2})$ with an additional factor $\log(n)$, is imposed in Condition \ref{as:sparsity}. Condition \ref{as:est_prop} requires certain convergence rates of both $\ell_1$ and $\ell_2$ norms of the estimators $\{\widehat{\bgamma}_k\}$ for the working PS models. These convergence rates are well-established in high-dimensional statistics and are satisfied by the Lasso estimator for generalized linear models \citep{van2008high} and the regularized calibration estimator \citep{tan2020regularized}. The convergence of the prediction error of $\pi_k^{(1)}(\bX,\widehat{\bgamma}_k)$ can be achieved through the convergence of $\widehat{\bgamma}_k$ combined with a Taylor expansion. This condition can be verified for working PS models under high-dimensional clustered data in Section \ref{sec:group}. The last part of Condition \ref{as:est_prop} becomes necessary when all the PS models are misspecified. In this case, we require that the probability limit $\blambda_{\ast}$ from (\ref{eq:EL-pi}) to be sparse. Condition \ref{as:sparsity} represents the regularity condition on the sparsity levels. In \cite{tan2020model} and \cite{ning2020robust}, the condition $\max\{s_1,s_2\} \log(m)/n^{1/2}=o(1)$ was assumed, which is slightly weaker than our Condition \ref{as:sparsity}. The additional term $(\min\{s_1,s_2\})^{1/2}$ in our condition arises from the feasibility of high-dimensional empirical likelihood \citep{chang2021high}. Moreover, the term $s_0^{1/2}$ is necessitated by the inclusion of the plug-in estimates $\pi_k^{(1)}(\bX, \widehat{\bgamma}_k)$ and $\pi_k(\bX, \widehat{\bgamma}_k)$ as covariates in the AOR defined in (\ref{eq:EL-outcome}), which introduces additional estimation errors from $\widehat{\bgamma}_k$. Nonetheless, incorporating $\pi_{k}^{(1)}(\bX, \widehat{\bgamma}_k)$ and $\pi_{k}(\bX, \widehat{\bgamma}_k)$ in the AOR is critical for achieving multiply robust inference under a misspecified OR model as demonstrated in (\ref{eq:expansion-PS})--(\ref{eq:outcome-bound-2}) in the Appendix. 

In the following, we first present the theoretical properties of the proposed empirical likelihood weights $\{\widetilde{p}_i: D_i = 1\}$ in \eqref{eq:EL-weights}.

\begin{proposition}\label{prop:EL}
Under Conditions \ref{as:ignorability}--\ref{as:est_prop}, assuming $w_{\ps} \asymp \{s_1\log(m)/n\}^{1/2}$ and $s_1\log(m)=o(n)$, the following results hold:

(i) if the PS model in (\ref{eq:MPS}) is correctly specified, then $(\widetilde{\lambda}_0, \widetilde{\blambda}_1^{\T}, \widetilde{\blambda}_2^{\T}) \stackrel{p}{\to} (\lambda_{0, \ast}, \blambda_{1, \ast}^{\T}, \blambda_{2, \ast}^{\T})$, $\|(\widetilde{\lambda}_0, \widetilde{\blambda}_1^{\T}) - (\lambda_{0, \ast}, \blambda_{1, \ast}^{\T})\|_2 = O_p\big(\{ s_1 \log(m) / n\}^{1/2}\big)$ and $\widetilde{\blambda}_2=\boldsymbol{0}$, with probability approaching $1$, where $\lambda_{0, \ast} = \E(D)\cdot \sum_{k = 1}^{q} d_k$, $\blambda_{1, \ast} = (d_1, \ldots, d_q)^{\T}$, and $\blambda_{2, \ast} = \bzero$; 

(ii) if the PS model in (\ref{eq:MPS}) is misspecified, the high-dimensional EL optimization problem in (\ref{eq:EL-observe})--(\ref{eq:EL-observe-calibration}) still admits a feasible solution with probability at least $1-\exp\{-c\log(m)\}$ for a constant $c>0$.
\end{proposition}

In the special case of (\ref{eq:MPS}), where $d_k = 1$ and $d_h = 0$ for $h \neq k$, indicating that the $k$th PS model is correctly specified, the first part of Proposition \ref{prop:EL} shows that the proposed EL estimator of $\pi(\bX)$ successfully selects the true propensity score model. Given the convergence of $\widehat{\bgamma}_k$, the estimated EL weights converge in probability to the true propensity scores, i.e., $\widetilde{\pi}_i\stackrel{p}{\to}\pi_i$. The second part of Proposition \ref{prop:EL} demonstrates that, even when all working PS models are misspecified, the empirical likelihood method still yields a solution that satisfies both the IBC constraints in \eqref{eq:EL-observe-IBC} and the soft covariate balancing constraints in \eqref{eq:EL-observe-calibration}. Consequently, despite the misspecification of all working PS models, the proposed weights retain covariate balancing properties and enable valid statistical inference.

We present the main theorem of the proposed method under three different settings.

\begin{theorem}
\label{thm:linear_main}
    Under Conditions \ref{as:ignorability}--\ref{as:sparsity}, $\omega_{\ps} \asymp \{s_1\log(m)/n\}^{1/2}$ and $\omega_{\oc} \asymp \{\log(m)/n\}^{1/2}$, 
    
    (i) if any of the working PS models $\{\pi_k(\bx, \bgamma_k)\}_{k = 1}^{q}$ or their linear combination in (\ref{eq:MPS}) is correctly specified, we have
    \begin{equation*}
    \begin{split}
        \widehat{\mu}_1 =& \frac{1}{n} \sum_{i = 1}^{n}\frac{D_i}{\pi_i} Y_i + \frac{1}{n} \sum_{i = 1}^{n} \frac{\pi_i-D_i}{\pi_i} \{\beta_{0, \ast} + \bbeta_{1, \ast}^{\T} \bb_{\ast}(\bX_i) + \bbeta_{2, \ast}^{\T} \bpi_{i, \ast}\}+ o_p(n^{-1/2});
    \end{split}
    \end{equation*}
    
    (ii) if the OR model in (\ref{eq:OR-model}) is correctly specified, we have
    \begin{equation*}
        \widehat{\mu}_1 = \frac{1}{n} \sum_{i = 1}^{n} {\E}(Y_i \vert \bX_i, D_i = 1)  + \frac{1}{n} \sum_{i = 1}^{n} \frac{D_i}{\pi_i^*} \{Y_i - {\E}(Y_i \vert \bX_i, D_i = 1)\} + o_p(n^{-1/2});
    \end{equation*}
    
    (iii) if both the OR model and at least one of the working PS models or their linear combination are correctly specified, we have
    \begin{equation*}
        \widehat{\mu}_1 = \frac{1}{n} \sum_{i = 1}^{n} {\E}(Y_i \vert \bX_i, D_i = 1) + \frac{1}{n} \sum_{i = 1}^{n} \frac{D_i}{\pi_i} \{Y_i - {\E}(Y_i \vert \bX_i, D_i = 1)\} + o_p(n^{-1/2}).
    \end{equation*}
\end{theorem}

Theorem \ref{thm:linear_main} shows that the influence functions of the proposed estimator $\widehat{\mu}_1$ have the same form under either the correctly specified PS model in (\ref{eq:MPS}) or the correctly specified OR model in (\ref{eq:OR-model}). The influence functions indicate the multiply robust inference property of the proposed method. When both the PS and OR models are correctly specified, the proposed estimator $\widehat{\mu}_1$ is locally efficient and attains the semiparametric efficiency bound \citep{hahn1998role}.

Based on Theorem \ref{thm:linear_main}, the variance of $\widehat{\mu}_1$ can be estimated by
\begin{equation}\label{eq:est_v}
   \widehat{V}_1=\frac{1}{n}\sum_{i=1}^{n}\bigg[\frac{D_i Y_i}{\widetilde{\pi}_i} -  \frac{D_i - \widetilde{\pi}_i}{\widetilde{\pi}_i} \{\widetilde{\beta}_0+\widetilde{\bbeta}_1^{\T} \widehat{\bb}(\bX_i)+\widetilde{\bbeta}_2^{\T} \widehat{\bpi}_i\}-\widehat{\mu}_1 \bigg]^2,
\end{equation}
and $\text{CI}_{1, \alpha} = \big( \widehat{\mu}_1-z_{1-\alpha/2}(\widehat{V}_1/n)^{1/2}, \widehat{\mu}_1+z_{1-\alpha/2}(\widehat{V}_1/n)^{1/2} \big)$ is the proposed $(1-\alpha)$-confidence interval of $\mu_1$, where $z_{1-\alpha/2}$ the upper $\alpha / 2$ quantile of the standard normal distribution. The confidence intervals of $\mu_0$ and $\tau$ can be constructed similarly. Proposition \ref{coro:linear_cover} shows the asymptotic coverage of the proposed confidence interval $\text{CI}_{1, \alpha}$.

\begin{proposition}\label{coro:linear_cover}
    Suppose the conditions in Theorem \ref{thm:linear_main} hold. Furthermore, assume the outcome regression error $\epsilon_1$ satisfies $1/c_1\leq\V(\epsilon_1|\bX)\leq c_1$ for $c_1>0$.
    Then, for $0<\alpha<1$, 
    $\lim_{n\to\infty}\PP(\mu_1\in \text{CI}_{1,\alpha} )=1-\alpha$.
\end{proposition}

In Proposition \ref{coro:linear_cover}, we require an additional assumption on the conditional variance of the error in the OR model. A similar assumption is made in \cite{ning2020robust} and \cite{tan2020model}. Under this assumption and conditions in Theorem \ref{thm:linear_main}, the variance of the influence function of the proposed estimator $\widehat{\mu}_1$ is bounded away from $0$ and $\infty$. The asymptotic normality of $\widehat{\mu}_1$ follows from Theorem \ref{thm:linear_main}.

\setcounter{equation}{0}
\section{ATE under Generalized Linear Models}
\label{sec:glm}

We extend the proposed multiply robust estimator in (\ref{eq:Est-mu-1}) to the setting where the outcome variable follows a generalized linear model. Let $f(Y(d)\mid\bX)$ denote the conditional density of the potential outcome $Y(d)$ given $\bX$. We assume the following OR model for $Y(d)$:
\begin{equation}\label{eq:GLM}
f(Y(d)\mid\bX)=h(Y(d),\phi)\exp\big([Y(d) \{\bbeta(d)^{\T} \bb_{\oc}(\bX)\} - g\{\bbeta(d)^{\T}\bb_{\oc}(\bX)\}] / a(\phi)\big)
\end{equation}
where $h(\cdot,\cdot)$, $a(\cdot)$ and $g(\cdot)$ are known functions, $\phi$ is the dispersion parameter, $\bbeta(d)$ is the unknown regression coefficient and $\bb_{\oc}(\bX)$ denotes a set of basis functions. Let $g^{\prime}(\cdot)$ and $g^{\prime\prime}(\cdot)$ denote the first- and second-order derivatives of $g(\cdot)$, respectively. Under model (\ref{eq:GLM}), the conditional variance of $Y(d)$ given $\bX$ is $\mathbb{V}(Y(d) \mid\bX) = g^{\prime\prime}\{\bbeta(d)^{\T}\bb_{\oc}(\bX)\} a(\phi)$. Accordingly, we balance the covariates $g^{\prime\prime}\{\bbeta(d)^{\T}\bb_{\oc}(\bX)\}\bb_{\oc}(\bX)$ when estimating $\mu_d$. In the following, we introduce the procedure to estimate $\mu_1$. The estimator of $\mu_0$ can be similarly constructed. 

First, obtain an initial estimator of $\bbeta(1)$ via regularized maximum likelihood estimation
\begin{equation*}
\widehat{\bbeta}_{ini} = \underset{\bbeta\in\mathbb{R}^{p_0+1}}{\argmin} \bigg( -\frac{1}{n}\sum_{i=1}^{n} D_i [Y \{\bbeta^{\T} \bb_{\oc}(\bX) \} - g\{\bbeta^{\T}\bb_{\oc}(\bX)\}] + w_{\oc}^{(0)}\|\bbeta\|_1\bigg),
\end{equation*}
where $w_{\oc}^{(0)}$ is a regularization parameter. 

Second, construct the EL weights $\{\widetilde{p}_i\}$ by solving the following optimization problem:
\begin{equation}
\label{eq:EL-lambda-equation-glm}
    \begin{split}
    \{\widetilde{p}_i:D_i=1\} = & \underset{p_i}{\operatorname{argmax}} \sum_{i=1}^{n}D_i\log(p_i),
    \text{ \ subject to \ } p_i \geq 0, \sum_{i=1}^{n}D_ip_i=1, \\
    & \sum_{i=1}^{n}D_ip_i\{\pi_k(\bX_i, \widehat{\bgamma}_k) - n_1/n\}=0 \text{ \ for $k = 1,\dots,q$ and } \\
    & \bigg| \sum_{i=1}^{n} D_ip_i \widetilde{b}_{ij}^{(g)} \bigg| \leq \omega_{\ps} \text{ \ for $j = 1,\dots,m$},
    \end{split}
\end{equation}
where $\widetilde{\bb}_i^{(g)} = (\widetilde{b}_{i0}^{(g)},\widetilde{b}_{i1}^{(g)}, \ldots, \widetilde{b}_{im}^{(g)})^{\T} = g^{\prime\prime} (\widehat{\bbeta}^{\T}_{ini}\bb_{\oc}(\bX_i))(1,\widehat{\bb}(\bX_i)^{\T})^{\T} - \overline{\bb}^{(g)}$, $\widehat{\bb}(\bX)$ is defined in \eqref{eq:basis}, and $\overline{\bb}^{(g)} = n^{-1}\sum_{i=1}^{n} g^{\prime\prime}(\widehat{\bbeta}^{\T}_{ini}\bb_{\oc}(\bX_i)) (1,\widehat{\bb}(\bX_i)^{\T})^{\T}$. Different from the EL formulation for the linear outcome model in (\ref{eq:EL-observe}), the soft calibration under GLMs is applied to the weighted basis functions $g^{\prime\prime} (\widehat{\bbeta}^{\T}_{ini}\bb_{\oc}(\bX_i)) (1, \widehat{\bb}(\bX_i)^{\T})^{\T}$, where the weights correspond to the conditional variance of $Y(1)$ given $\bX$ under the working GLM in (\ref{eq:GLM}). The high-dimensional EL optimization in (\ref{eq:EL-lambda-equation-glm}) can be solved via a regularized unconstrained optimization similar to (\ref{eq:EL-pi}). 

Third, fit the AOR by minimizing the regularized objective function
\begin{equation*}
\begin{split}
\{\widetilde{\beta}_0,\widetilde{\bbeta}_1, \widetilde{\bbeta}_2\} = \underset{\beta_0, \bbeta_1, \bbeta_2}{\argmin} \bigg(& \frac{1}{n}\sum_{i=1}^{n} \frac{D_i}{\widetilde{\pi}_i^2} [ g\{\beta_0+\bbeta_1^{\T}\widehat{\bb}(\bX_i)+\bbeta^{\T}_2\boldsymbol{\widehat{\pi}}_i\} \\
& - Y_i\{\beta_0+\bbeta_1^{\T}\widehat{\bb}(\bX_i)+\bbeta^{\T}_2\boldsymbol{\widehat{\pi}}_i\} ] + w_{\oc}(\|\bbeta_1\|_1+\|\bbeta_2\|_1) \bigg),
\end{split}
\end{equation*}
where $\widetilde{\pi}_i=(n\widetilde{p}_i)^{-1}$ is obtained from (\ref{eq:EL-lambda-equation-glm}) and $w_{\oc}$ is the penalty parameter. Finally, the estimator $\widehat{\mu}_1$ is constructed as 
\begin{equation}
\label{eq:linear_aipw}
    \widehat{\mu}_1=\frac{1}{n}\sum_{i=1}^{n}\frac{D_iY_i}{\widetilde{\pi}_i}-\frac{1}{n}\sum_{i=1}^{n}\frac{D_i-\widetilde{\pi}_i}{\widetilde{\pi}_i}\{g^{\prime}(\widetilde{\beta}_0+\widetilde{\bbeta}^{\T}_1\widehat{\bb}(\bX_i)+\widetilde{\bbeta}_2^{\T}\boldsymbol{\widehat{\pi}}_i)\}.
\end{equation}
Compared to the procedure in Section \ref{sec:lm_method} under the linear OR model, the proposed method under GLMs requires only an initial estimate of $\bbeta(d)$ under the working OR model specified in (\ref{eq:GLM}). Following the same procedure, we can construct the estimate $\widehat{\mu}_0$ of $\mu_0$, and define the estimated ATE as $\widehat{\tau} = \widehat{\mu}_1 - \widehat{\mu}_0$.

The $(1-\alpha)$-confidence interval of $\mu_1$ is constructed as $\text{CI}_{1, \alpha}=\big( \widehat{\mu}_1-z_{1-\alpha/2}(\widehat{V}_1/n)^{1/2}, \widehat{\mu}_1+z_{1-\alpha/2}(\widehat{V}_1/n)^{1/2} \big)$, where 
$$\widehat{V}_1 = \frac{1}{n} \sum_{i=1}^{n}\bigg[ \frac{D_i Y_i}{\widetilde{\pi}_i} - \frac{D_i - \widetilde{\pi}_i}{\widetilde{\pi}_i} g^{\prime}\{\widetilde{\beta}_0 + \widetilde{\bbeta}_1^{\T} \widehat{\bb}(\bX_i) + \widetilde{\bbeta}_2^{\T} \widehat{\bpi}_i\}-\widehat{\mu}_1 \bigg]^2$$
is the estimated variance of $\widehat{\mu}_1$. 
Similar to Theorem \ref{thm:linear_main}, the theoretical results of the proposed estimator and the confidence interval under the GLM setting are provided in the SM.

\setcounter{equation}{0}
\section{ATE under Clustered Data}
\label{sec:group}

In this section, we assume that the population distribution $\mathcal{G}$ is from a mixture of $v$ subgroups. Let $\bU_i = (U_{i1}, \ldots, U_{iv})^{\T}$ be the unobserved subgroup indicator for the $i$th observation, where $U_{i \ell} = 1$ and $U_{i h} = 0$ for $h \neq \ell$ if the $i$th observation belongs to the $\ell$th cluster. Let $\bmu_{x,\ell}$ be the mean of $\bX_i$ for the $\ell$th cluster, and $\bM = (\bmu_{x,1}, \ldots, \bmu_{x,v})^{\T}$. We assume the following latent structure:
\be\label{eq:cluster-model}
(\bX_1, \ldots, \bX_n)^{\T} = \bU \bM + \bepsilonb,
\ee 
where $\bU = (\bU_1, \ldots, \bU_n)^{\T}$, $\bepsilonb = (\bepsilonb_1, \ldots, \bepsilonb_n)^{\T}$, and $\cov(\bepsilonb_i) = \bSigma_x$. Let $\pi(\bX_i, \ell) = \PP(D_i = 1 \mid \bX_i, U_{i \ell} = 1)$ be the propensity score of the $i$th observation, given it is from the $\ell$th subgroup. If $\bU_i \not\perp (D_i, \bX_i)$, then the subgroups exhibit heterogeneity in both the distribution of covariates $\bX$ and the treatment assignment mechanism, resulting in differing propensity scores across clusters.

In the following, we discuss the estimation of the propensity score and show it satisfies Condition \ref{as:est_prop} under clustered data. Under the model in (\ref{eq:cluster-model}), \cite{Jin2017, Abbe2022} showed that spectral clustering can consistently recover the true class labels under a signal-to-noise ratio condition, where signal and noise are determined by the Euclidean distance among the cluster means $\{\bmu_{x,1}, \ldots, \bmu_{x,v}\}$ and the covariance matrix $\bSigma_{x}$, respectively. As discussed in Section \ref{sec:pre}, multiple PS models can be fitted based on the estimated cluster labels, followed by the application of the proposed multiply robust inference procedure for estimating the ATE. Note that different clustering methods and multiple candidate values for the number of clusters could also be implemented.

Given a set of candidate PS models $\{\tilde{\pi}_h(\bX, \bgamma_{h}): h = 1, \ldots, q_c\}$ for the subgroups and the estimated cluster labels $\widehat{\bU}_i = (\widehat{U}_{i1}, \ldots, \widehat{U}_{iv})^{\T}$ based on $\bX_1, \ldots, \bX_n$, we proceed as follows. For notational simplicity, we consider a single set of estimated clusters, though alternative clustering methods can also be implemented. Let $\widehat{\bgamma}_{h,\ell}$ be the estimate of the parameter $\bgamma_{h}$ in the $h$th PS model, based on the observations from the $\ell$th estimated cluster with $\widehat{U}_{i \ell} = 1$. The estimated working PS model for the whole population is defined as
\be
\pi_{h_1 \ldots h_v}(\bX_i, \widehat{\bgamma}_{h_1 \ldots h_v}) = \sum_{\ell = 1}^{v} \tilde{\pi}_{h_\ell}(\bX_i, \widehat{\bgamma}_{h_{\ell},\ell}) 1(\widehat{U}_{i \ell} = 1)
\label{eq:est-PS-cluster}\ee
for $i = 1, \ldots, n$, and $h_1, \ldots, h_v = 1, \ldots, q_c$, where $\widehat{\bgamma}_{h_1 \ldots h_v}^{\T}=(\widehat{\bgamma}_{h_1,1}^{\T}, \ldots, \widehat{\bgamma}_{h_v,v}^{\T})$ and $1(\cdot)$ denotes the indicator function. The total number of working PS models for the whole population is $q_c^v$. To study asymptotic properties of $\pi_{h_1 \ldots h_v}(\bX_i, \widehat{\bgamma}_{h_1 \ldots h_v})$ in (\ref{eq:est-PS-cluster}) and verify Condition \ref{as:est_prop}, we make the following conditions on the clustering method.

\begin{condition}\label{as:group-size}
There exists a positive constant $c_{\min}$ such that $\mathbb{P}_{\mathcal{G}}(U_{i \ell} = 1) > c_{\min}$ for $\ell = 1, \ldots, v$, where $\mathbb{P}_{\mathcal{G}}(\cdot)$ denotes the probability under the mixture distribution $\mathcal{G}$. Furthermore, the clustering error rate satisfies $ n^{-1} \sum_{i=1}^{n}1(\bU_{i} \neq \widehat{\bU}_{i}) = O\{s_1\log(p) / n\}$.
\end{condition}

Condition \ref{as:group-size} assumes that each cluster has a non-trivial probability under the population distribution $\mathcal{G}$. Since separate PS models are fitted within each cluster, the condition ensures a sufficient number of observations for each model. Additionally, Condition \ref{as:group-size} requires that the estimated clusters converge at a specific rate, $O_p\{s_1 \log(p) / n\}$, where $s_1$ is the sparsity level of the parameter in the PS models. This condition can be satisfied by spectral clustering if there is sufficient separation among the cluster means $\{\bmu_{x,1}, \ldots, \bmu_{x,v}\}$ \citep{Jin2017, Abbe2022}. Particularly, \cite{cai2019chime} showed that $O_p\{s_3 \log(p) / n\}$ is the minimax convergence rate for two-group clustering with sparse signals in the means, where $s_3$ represents the sparsity level of the clustering signals. 

The following theorem establishes the convergence of estimated PS models in \eqref{eq:est-PS-cluster} for clustered data under logistic regression within each subgroup. In particular, conditions of Theorem \ref{thm:linear_main}, which are required for multiply robust inference, are satisfied.

\begin{theorem}
\label{thm:group_main}
Suppose that Condition \ref{as:group-size} holds, and $\bX$ follows a sub-Gaussian distribution. For $\ell=1,\dots,v$, consider the logistic regression model $\tilde{\pi}_{h_\ell}(\bX_i, \bgamma_{h_{\ell},\ell}) = \{ 1 + \exp(-\bX^{\T} \bgamma_{h_{\ell},\ell}) \}^{-1}$, and the estimator $\widehat{\bgamma}_{h_{\ell},\ell}$ is obtained via $\ell_1$-regularized logistic regression with a penalty of order $\{\log(p)/n\}^{1/2}$, using the data $\{(\bX_i,D_i): \widehat{U}_{il}=1\}$ from the estimated subgroup. If the true parameter $\bgamma^*_{h_{\ell},\ell}$, defined in (S5.11) of the SM, satisfies $\|\bgamma^*_{h_{\ell},\ell}\|_0\leq s_1$ and $c_0\leq\tilde{\pi}_{h_\ell}(\bX, \bgamma_{h_{\ell},\ell}^*)\leq 1-c_0$ for a positive constant $c_0\in(0,1/2)$, then the PS model defined in \eqref{eq:est-PS-cluster} satisfies the conditions required by Theorem \ref{thm:linear_main}.
\end{theorem}

By Theorem \ref{thm:group_main}, the conclusions of Theorem \ref{thm:linear_main} apply to the proposed procedure using the estimated PS model $\{ \pi_{h_1 \ldots h_v}(\bX_i, \widehat{\bgamma}_{h_1 \ldots h_v}) \}$ in (\ref{eq:est-PS-cluster}), obtained from a clustering algorithm for high-dimensional clustered data.

\setcounter{equation}{0}
\section{Simulation}
\label{sec:num}

In this section, we evaluate the performance of the proposed multiply robust inference procedure—debiased high-dimensional empirical likelihood weighting (DELW)—through simulation studies, and compare it with five existing doubly robust inference methods. The results under GLM settings are reported in the SM.

\subsection{Multiple Robustness under non-Clustered Data}

We first evaluate the proposed method DELW using multiple PS models under non-clustered data. 
The covariates $\bX$ are generated from the normal distribution with mean $\boldsymbol{0}$ and covariance $\bSigma_x = (\sigma_{x,j_1j_2})$, where $\sigma_{x,j_1j_2}=0.5^{|j_1 - j_2|}$. The treatment indicator $D$ is generated from the logistic regression under two designs of propensity scores,
\bea
& \mbox{PS1:} & \mathbb{P}(D = 1 \mid \bX) = \big[ 1 + \exp\{-X_{1} + X_{2}/2 - X_{3}/4 - (X_{4} + X_{5} - X_{6}) / 10\} \big]^{-1}, \nn \\ 
& \mbox{PS2:} & \mathbb{P}(D = 1 \mid \bX) = \big[ 1 + \exp\{-Z_{1} + Z_{2}/2 - Z_{3}/4 - (Z_{4} + Z_{5} - Z_{6}) / 10\} \big]^{-1}, \nn
\eea
where $(Z_1, \ldots, Z_p)^{\T} = \bZ(\bX) = (\exp(X_1/2),X_2/\{1+\exp(X_1)\}+10,(X_1X_3/25+0.6)^3,(X_2+X_4+20)^2,X_6,\exp(X_6+X_7),X_9^2,X_7^3-20,X_9,\dots,X_p)^{\T}$ is a nonlinear transformations of $\bX$ with the first 8 covariates in $\bX$ being replaced. 
Two designs of data generation, OR1 and OR2, are applied to the potential outcomes, where
\bea
&\mbox{OR1:}&
Y(0) = 1 + 0.291 \sum_{j = 5}^{10} X_j 
+\epsilon_{0} \mbox{ \ and \ }
Y(1) = 2 + 0.137 \sum_{j = 5}^{8} X_j
+\epsilon_{1}, \nn \\
&\mbox{OR2:}& Y(0) = 1 + 0.291 \sum_{j = 5}^{10} Z_j +\epsilon_{0} \mbox{ \ and \ } Y(1) = 2 + 0.137 \sum_{j = 5}^{8} Z_j +\epsilon_{1}, \nn
\eea
and $\epsilon_{0}, \epsilon_{1}$ are generated from $N(0, 1)$, independent of $\bX$. The observed response is obtained via $Y = D Y(1) + (1 - D) Y(0)$. These designs provide four data generation mechanisms for $(\bX,Y,D)$. Under each setting, a sample of size $n$ is generated IID. The parameter of interest is the ATE $\tau = \mathbb{E}\{Y(1)-Y(0)\}$. We set $n = 500, 1000$ and $p = 1000, 2000$. Each setting of the simulation studies is repeated $500$ times.

We compare the proposed method (DELW) with the high-dimensional covariate balancing propensity score (CBPS) method in \cite{ning2020robust}, model-assisted regularized calibration (RC) method in \cite{tan2020model}, approximate residual balancing (RB) method in \cite{athey2018approximate}, the double selection (DS) method in \cite{belloni2014inference}, and the regularized AIPW (rAIPW) method, which naively incorporates regularized estimates of PS and OR models into the AIPW formulation. All five existing methods utilize a single linear OR model and a single logistical PS model with covariates $\bX$. The four designs of data generation result in four combinations of model specifications for the doubly robust methods: PS1 and OR1 (both models correct), PS1 and OR2 (only PS model correct), PS2 and OR1 (only OR model correct), PS2 and OR2 (both models incorrect). The proposed method DELW uses the linear regression model with covariate $\bX$ for OR and two working logistical regression models for PS, one with covariates $\bX$ and the other with covariates $\bZ$. For the proposed method, as one of the working PS models is correctly specified under each of the four designs, the linear combination condition for PS in (\ref{eq:MPS}) is satisfied. However, the outcome model is still misspecified under the design OR2.

Both OR and PS models in all the methods are estimated using $\ell_1$-regularized regression. We use the 5-fold cross-validation to determine the penalty parameters in DELW and rAIPW. We use R packages \texttt{CBPS}, \texttt{balanceHD}, and \texttt{hdm} to implement CBPS and RC, RB, and DS, respectively. We use the default values of the tuning parameters in the R packages for those four methods.

The results under $n = 500$ are presented in Tables \ref{tab:linear_500_correct_pro} and \ref{tab:linear_500_mis_pro} for the PS1 and PS2 designs, respectively. The results under $n = 1000$ are reported in the SM. From the two tables, we observe that the proposed method DELW has a small bias and an accurate coverage rate close to the nominal level of 95\% for the confidence interval under all scenarios. We compare its performance with the existing methods under the four data generation mechanisms. Under the design of PS1 and OR1, all methods demonstrate reasonable coverage rates and relatively small bias. Both PS and OR models are correctly specified for all methods. Meanwhile, the RMSE of the proposed method, which allows for an additional estimated PS model, is comparable to that of rAIPW and smaller than the RMSEs of all other methods. Similarly, their performance is comparable under the design of PS2 and OR1, as the outcome model is correctly specified for all methods. However, the proposed method still has a relatively smaller RMSE than most existing methods. Under the design of PS1 and OR2, all methods have a correctly specified PS model but a misspecified outcome model. The proposed method is much more efficient than all other methods, with the smallest standard error and shortest confidence interval. The efficiency gain is attributed to adding additional variables in the soft covariate balancing and augmented outcome regression. Under the design of PS2 and OR2, both the PS and OR models are misspecified for the existing methods. Those methods display significantly larger bias and higher standard error and RMSE, resulting in much wider confidence intervals and coverage rates below the nominal level. In contrast, the proposed method DELW, by incorporating an additional propensity model, exhibits the smallest RMSE and achieves robustness inference with coverage close to the nominal level. In summary, the proposed estimator shows a smaller RMSE, enhanced robustness to model misspecification, and an accurate coverage rate of the confidence interval. Additional simulation studies under GLM for the response variable are provided in the SM.

\subsection{Clustered Data}

Then, we evaluate the performance of the proposed DELW method under clustered data by simulation. We consider the data generation mechanisms, where the covariates $\{\bX_i\}_{i=1}^{n}$ are independently drawn from two distinct clusters. The first half of samples ($i=1,\dots,n/2$) are from cluster 1, where $\bX_i$ is generated from the multivariate normal distribution with mean $-\boldsymbol{\mu}_x$ and covariance $\bSigma_x = (\sigma_{x,j_1j_2})$ with $\sigma_{x,j_1j_2}=0.5^{|j_1-j_2|}$. The second half of the samples ($i=n/2+1,\dots,n$) are from cluster 2, where $\bX_i$ is generated from the multivariate normal distribution with mean $\boldsymbol{\mu}_x$ and the same covariance $\bSigma_x$. The discriminant vector $\boldsymbol{\mu}_x=(\mu_{x,1}, \ldots, \mu_{x,p})^{\T}$ is sparse with only the second set of $10$ entries being nonzero, where $\mu_{x,j} = 1$ for $j = 11, \ldots, 20$ and $\mu_{x,j} = 0$ for all other $j$. This setup is similar to that considered by \cite{cai2019chime} for high-dimensional clustering. 

Two different designs of the propensity scores are considered for the two clusters, where $\mathbb{P}(D = 1 \mid \bX)$ follows PS1 for cluster 1 and PS1 with $\bX$ replaced by $-\bX$ for cluster 2, or equivalently, $\mathbb{P}(D = 0 \mid \bX)$ follows PS1 for cluster 2. For the proposed method DELW, we apply the high-dimensional clustering algorithm based on the covariate $\bX$ by the R package \texttt{HDclassif} with two clusters and fit $\ell_1$ regularized logistic regression on the two estimated clusters separately. The clustering algorithm employs the expectation-maximization (EM) algorithm to estimate the model parameters, which are sensitive to the initial values of the parameters. We use two sets of initial values to generate two propensity score models, both incorporated into the proposed procedure. All OR and PS models are estimated by $\ell_1$ regularized regression. Each setting of the simulation studies is repeated $500$ times.

The results are presented in Table \ref{tab:cluster_1000}. From this table, we observe that the proposed method DELW has a small bias and an accurate coverage rate close to the nominal level of 95\% for the confidence interval under both designs for the potential outcomes. Under the OR1 design, where the outcome regression model is correctly specified for all methods, the results of those methods are comparable, achieving reasonable coverage with a small bias. However, the proposed method DELW has the smallest RMSE and the shortest confidence interval. Under the OR2 design, where the outcome regression model is misspecified for all methods, the five existing methods display significant biases, large standard errors, and inaccurate coverage rates below the nominal level. In contrast, DELW also shows a much smaller RMSE and a shorter confidence interval with a coverage rate closer to the nominal level, indicating the validity of the proposed method under clustered data. 

\begin{table}[!ht]
\centering
\caption{Bias, standard error (Std), root-mean-squared error (RMSE), coverage probability of $95\%$ confidence intervals (coverage), and length of $95\%$ confidence intervals (CI width) of the proposed method DELW and five existing methods for the average treatment effect under $n = 500$ and the PS1 design for propensity scores.}
\footnotesize
\begin{tabular}{c|c|cccccc}
  \hline
 & & DELW & CBPS & RC & RB & rAIPW & DS \\ 
  \hline
$p$&Measures& \multicolumn{6}{c}{OR1 design for outcome regression} \\ 
  \hline
\multirow{5}{2em}{1000}&Bias & 0.0054 & 0.0017 & -0.0130 & 0.0033 & -0.0022 & -0.0025 \\ 
&Std& 0.0967 & 0.0981 & 0.1002 & 0.1096 & 0.0964 & 0.1242 \\ 
&RMSE & 0.0968 & 0.0982 & 0.1011 & 0.1097 & 0.0965 & 0.1243 \\ 
&Coverage & 0.9500 & 0.9450 & 0.9200 & 0.9600 & 0.9450 & 0.9550 \\ 
&CI width & 0.3792 & 0.3873 & 0.3659 & 0.4318 & 0.3663 & 0.5182 \\ 
  \hline
\multirow{5}{2em}{2000}&Bias & 0.0039 & 0.0035 & -0.0140 & 0.0042 & -0.0059 & -0.0061 \\ 
&Std & 0.0991 & 0.1009 & 0.1009 & 0.1099 & 0.0964 & 0.1402 \\ 
&RMSE & 0.0991 & 0.1010 & 0.1020 & 0.1100 & 0.0965 & 0.1404 \\ 
&Coverage & 0.9300 & 0.9350 & 0.8900 & 0.9500 & 0.9200 & 0.9700 \\ 
&CI width & 0.3761 & 0.3871 & 0.3655 & 0.4278 & 0.3650 & 0.5746 \\ 
   \hline
 & & \multicolumn{6}{c}{OR2 design for outcome regression} \\ 
  \hline
\multirow{5}{2em}{1000}&Bias & 0.0094 & 0.0176 & 0.0191 & 0.0512 & 0.0191 & 0.0543 \\ 
&Std & 0.2027 & 0.4258 & 0.4350 & 0.5148 & 0.4511 & 0.4913 \\ 
&RMSE & 0.2029 & 0.4262 & 0.4355 & 0.5173 & 0.4514 & 0.4942 \\ 
&Coverage & 0.9200 & 0.9350 & 0.9200 & 0.9400 & 0.9050 & 0.9600 \\ 
& CI width & 0.6830 & 1.4209 & 1.3134 & 1.7057 & 1.3140 & 1.6734 \\
\hline 
\multirow{5}{2em}{2000}&  Bias & 0.0105 & -0.0246 & -0.0167 & -0.0008 & -0.0334 & -0.1299 \\ 
&Std & 0.2239 & 0.4193 & 0.4497 & 0.6513 & 0.5167 & 1.0984 \\ 
&RMSE & 0.2241 & 0.4201 & 0.4500 & 0.6513 & 0.5178 & 1.1061 \\ 
&Coverage & 0.9050 & 0.9150 & 0.8800 & 0.9300 & 0.8850 & 0.9750 \\ 
&CI width & 0.6714 & 1.3366 & 1.3042 & 1.7164 & 1.3058 & 2.2246 \\
\hline  
\end{tabular}
\label{tab:linear_500_correct_pro}
\end{table}

\begin{table}[!ht]
\centering
\caption{Bias, standard error (Std), root-mean-squared error (RMSE), coverage probability of $95\%$ confidence intervals (coverage), and length of $95\%$ confidence intervals (CI width) of the proposed method DELW and five existing methods for the average treatment effect under $n = 500$ and the PS2 design for propensity scores.}
\footnotesize
\begin{tabular}{c|c|cccccc}
  \hline
 & & DELW & CBPS  & RC & RB &rAIPW & DS \\ 
  \hline
$p$&Measures& \multicolumn{6}{c}{OR1 design for outcome regression} \\ 
  \hline
\multirow{5}{2em}{1000}&Bias & -0.0111 & -0.0060 & -0.0149 & -0.0072 & -0.0092 & -0.0066 \\ 
  &Std & 0.0976 & 0.0961 & 0.1019 & 0.1064 & 0.0980 & 0.1230 \\ 
  &RMSE & 0.0980 & 0.0961 & 0.1027 & 0.1064 & 0.0981 & 0.1228 \\ 
  &Coverage & 0.9400 & 0.9600 & 0.9200 & 0.9500 & 0.9350 & 0.9400 \\ 
  &CI width & 0.3760 & 0.3854 & 0.3692 & 0.4253 & 0.3692 & 0.4959 \\ \hline
  \multirow{5}{2em}{2000}&Bias & -0.0135 & -0.0103 & -0.0157 & -0.0142 & -0.0129 & -0.0270 \\ 
  &Std & 0.1003 & 0.1012 & 0.1016 & 0.1099 & 0.1006 & 0.1300 \\ 
  &RMSE & 0.1010 & 0.1015 & 0.1026 & 0.1106 & 0.1012 & 0.1324 \\ 
  &Coverage & 0.9250 & 0.9400 & 0.9250 & 0.9250 & 0.9200 & 0.9600 \\ 
  &CI width & 0.3734 & 0.3825 & 0.3669 & 0.4182 & 0.3673 & 0.5445 \\ \hline
 & & \multicolumn{6}{c}{OR2 design for outcome regression} \\ 
  \hline
\multirow{5}{2em}{1000}&Bias & 0.0133 & 0.2239 & 0.2056 & 0.1925 & 0.2220 & 0.1952 \\ 
&Std & 0.1832 & 0.3637 & 0.3529 & 0.3281 & 0.3265 & 0.3952 \\ 
&RMSE & 0.1832 & 0.4263 & 0.4077 & 0.3797 & 0.3942 & 0.4398 \\ 
&Coverage & 0.9300 & 0.8200 & 0.7850 & 0.8300 & 0.7850 & 0.9200 \\ 
&CI width & 0.6448 & 1.1885 & 1.0534 & 1.2984 & 1.0554 & 1.4325 \\ \hline
\multirow{5}{2em}{2000}&  Bias & 0.0024 & 0.2197 & 0.1949 & 0.1888 & 0.2033 & 0.1916 \\ 
&Std & 0.1701 & 0.3069 & 0.3084 & 0.3322 & 0.2957 & 0.3797 \\ 
&RMSE & 0.1697 & 0.3768 & 0.3642 & 0.3814 & 0.3582 & 0.4245 \\ 
&Coverage & 0.9150 & 0.7900 & 0.7600 & 0.8800 & 0.7650 & 0.9450 \\ 
&CI width & 0.6220 & 1.1491 & 1.0182 & 1.2987 & 1.0190 & 1.5172 \\ \hline  
\end{tabular}
\label{tab:linear_500_mis_pro}
\end{table}

\begin{table}[!ht]
\centering
\caption{Bias, standard error (Std), root-mean-squared error (RMSE), coverage probability of $95\%$ confidence intervals (coverage), and length of $95\%$ confidence intervals (CI width) of the proposed method DELW and five existing methods for the average treatment effect under the clustered data with sample size $n = 1000$ and dimension $p = 1000$.}
\small
\begin{tabular}{c|cccccc}
  \hline
 & DELW & CBPS  & RC & RB &rAIPW & DS \\ 
  \hline
Measures & \multicolumn{6}{c}{OR1 design for outcome model} \\
\hline
Bias & 0.0049 & 0.0028 & 0.0025 & 0.0013 & 0.0052 & -0.0035 \\ 
  Std & 0.0706 & 0.0721 & 0.0709 & 0.0740 & 0.0711 & 0.0725 \\ 
  RMSE & 0.0708 & 0.0721 & 0.0709 & 0.0740 & 0.0713 & 0.0726 \\ 
  Coverage & 0.9400 & 0.9300 & 0.9350 & 0.9550 & 0.9400 & 0.9550 \\ 
  CI width & 0.2705 & 0.2746 & 0.2856 & 0.2873 & 0.2858 & 0.2847 \\ \hline 
 & \multicolumn{6}{c}{OR2 design for outcome model} \\
\hline
Bias & 0.0113 & 0.0204 & 0.0353 & 0.0402 & 0.0357 & 0.0182\\ 
Std & 0.1172 & 0.2950 & 0.2927 & 0.3358 & 0.3050 & 0.3491 \\ 
RMSE & 0.1178 & 0.2957 & 0.2948 & 0.3382 & 0.3071 & 0.3496 \\ 
Coverage & 0.9500 & 0.9150 & 0.8900 & 0.8950 & 0.9000 & 0.9200\\ 
CI width &  0.4505 & 0.9596 & 0.9565 & 1.0913 & 0.9569 & 0.9838\\ \hline
\end{tabular}
\label{tab:cluster_1000}
\end{table}

\setcounter{equation}{0}
\section{Real Data Analysis}
\label{sec:realdata}

This section presents a real-world data example from heterogeneous sources to demonstrate the need for developing a multiply robust method for estimating the ATE. The right heart catheterization dataset was initially reported in \cite{connors1996effectiveness} to study the treatment effect of RHC for seriously ill patients in the intensive care unit (ICU) on their subsequent one-month survival rate. The data set contains $5735$ patients with $72$ features from two studies across five medical centers. In the original work, \cite{connors1996effectiveness} combined the data sets from those two potentially heterogeneous studies and used a single logistic regression to fit the propensity scores. The following works also took the similar strategy, for example, \cite{hirano2001estimation, vermeulen2015bias, tan2020model}. However, the mechanism of treatment selection could be different between the two studies and the five medical centers, and the subgroup indicators were not released in the dataset. 

\begin{figure}[!t]
    \centering
    \begin{subfigure}{0.46\textwidth}
    \centering
    \includegraphics[width=\textwidth]{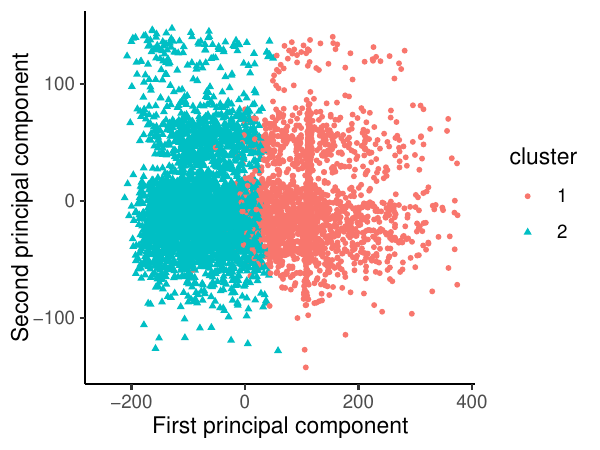}
    \caption{Estimated clusters}
  \end{subfigure}
  \hfill
  \begin{subfigure}{0.46\textwidth}
    \centering
    \includegraphics[width=\textwidth]{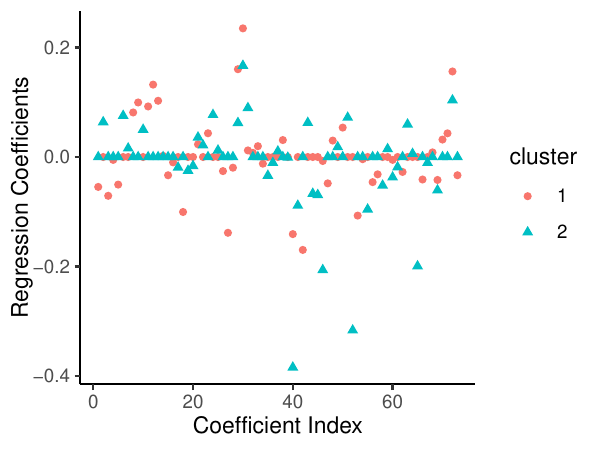}
    \caption{Estimated PS coefficients}
  \end{subfigure}
    \caption{Scatter plot of the first two principal components for the observations under the estimated clusters based on their covariates (left panel); the corresponding estimated regression coefficients of the PS models under the two estimated clusters (right panel).}
    \label{fig:ill}
\end{figure}

For each patient, the treatment indicator $D$ takes the value $1$ if the patient received the RHC treatment within 24 hours of admission to the ICU, and $D=0$ otherwise. The outcome variable $Y$ takes the value $1$ if the patient survived up to $30$ days, and $Y=0$ otherwise. The parameter of interest is the ATE of applying RHC on the survival probability. We apply two clustering algorithms by R packages \texttt{HDclassif} and \texttt{ClusterR} with two clusters based on the covariates. We estimate PS by fitting logistic regression models of $D$ on the estimated clusters separately. The results of the first clustering algorithm are presented in Figure \ref{fig:ill}, which shows that the estimated clusters are separated along the direction of the first principal component. The right panel of Figure \ref{fig:ill} shows that the estimated regression coefficients of the PS models differ a lot between the two clusters, indicating the heterogeneity of the propensity scores. The estimated clusters by the second algorithm are presented in the SM.

We use a logistic regression model to fit the binary outcome $Y$ and construct the covariate vector $\bX$ with a dimension $p=498$, which includes all main effects and two-way interactions of the $72$ original covariates, excluding those with a fraction of nonzero values lower than $10\%$ of the sample size. Table \ref{tab:rhc} reports the estimates of the one-month survival probabilities in the treatment and control group, along with the estimated ATE obtained by the proposed DELW, RC, and rAIPW methods, where the results for RC and rAIPW are taken from Table 2 in \cite{tan2020model}. The estimated standard errors of the three methods are similar. According to the results from RC and rAIPW, the ATE is significantly negative, suggesting that taking the RHC treatment would significantly lower the one-month survival probability. The conclusion raises practical concerns. However, the estimated ATE by the proposed method is smaller in absolute value and not significantly different from zero at the $0.05$ significance level.

From a medical perspective, the effectiveness of the RHC test is controversial. \cite{stevenson2005evaluation} concluded that adding RHC to careful clinical assessment increased anticipated adverse events but did not affect overall mortality and hospitalization. \cite{GARAN2020903} showed that using RHC is associated with improved survival from cardiogenic shock. Currently, RHC is still recommended in clinical practice \citep{fleisher20142014}. To evaluate certain critical cardiopulmonary conditions, RHC is performed when it is believed to be crucial for adjusting treatment plans and when it is unlikely to exacerbate the existing condition. Our analyses of the RHC dataset indicate that RHC treatment had an insignificant impact on early hospital mortality. The finding reflects a positive view on the use of RHC, aligning with its current widespread application.

\begin{table}[!t]
    \caption{Estimates $\pm \ 2 \ \times$ standard errors of one-month survival probabilities (in treatment and control groups) and the ATE by the proposed DELW, RC, and rAIPW methods.}
    \centering\small
    \begin{tabular}{c|c|c|c}
    \hline
        Method & DELW & RC & rAIPW\\
        \hline    $\mu_1$ &0.655$\pm$0.022&0.635$\pm$0.021&0.636$\pm$0.021\\      $\mu_0$ &0.678$\pm$0.016&0.688$\pm$0.016&0.691$\pm$0.016\\
        ATE $\tau$ &-0.022$\pm$0.026&-0.053$\pm$0.025&-0.055$\pm$0.025\\
       \hline
    \end{tabular}
    \label{tab:rhc}
\end{table}

\setcounter{equation}{0}
\section{Discussion}
\label{sec:con}

This paper proposes a new method for multiply robust inference of ATEs under high-dimensional covariates. The method is particularly useful for clustered data with heterogeneous treatment assignment mechanisms, where correctly specifying a single parametric PS model could be difficult. The proposed procedure complements the high-dimensional doubly robust procedures by utilizing multiple PS models. Although machine learning and nonparametric estimation of propensity scores and outcome regression can be applied to estimate ATEs, these methods generally do not incorporate multiple estimated models, and their properties under general high-dimensional settings remain unclear. The proposed method provides valid statistical inference of the ATE, if any working PS models, their linear combination, or the OR model is correctly specified. The confidence interval can be computed efficiently without requiring bootstrap procedures. In the presence of unobserved clusters, the ATE can be identified under Condition \ref{as:ignorability} of ignorability, which is an easier case than causal inference under unobserved confounders. An important future direction is to extend multiply robust inference to settings with unobserved confounders where the ignorability assumption doesn't hold.

\bibliographystyle{apalike}
\bibliography{reference.bib}

\newpage

\appendix

\section{Multiple Robustness by Augmented Calibration and Outcome Regression}



In the following, we outline the key steps to explain the importance of soft calibration using augmented covariates with PS derivatives, and augmented outcome regression to achieve multiply robust inferences under high-dimensional covariates. Recall that the subscript $\ast$ denotes the probability limit of an estimate under either a correctly specified or misspecified model. Calibrating the PS derivatives $\{\pi_{k}^{(1)}(\bX, \widehat{\bgamma}_k)\}_{k=1}^{q}$ leads to the approximation: 
\be\widehat{\mu}_1 \approx \frac{1}{n}\sum_{i = 1}^{n} \frac{D_i Y_i}{\widetilde{\pi}_i} - \frac{1}{n} \sum_{i = 1}^{n} \frac{D_i - \widetilde{\pi}_i}{\widetilde{\pi}_i} \{\beta_{0, \ast}+\bbeta_{1, \ast}^{\T} \widehat{\bb}(\bX_i) + \bbeta_{2, \ast}^{\T} \widehat{\bpi}_i\},\label{eq:expansion}\ee
where the regression coefficients from the AOR in (\ref{eq:EL-outcome}) are replaced by their probability limits. 

Suppose the OR model in (\ref{eq:OR-model}) is correctly specified, then $\bbeta_{1k, \ast} = \bzero$ for $k = 1, \ldots, q$, $\bbeta_{2, \ast} = \bzero$. In this case, $\widehat{\mu}_1$ can be asymptotically expressed as 
\be\widehat{\mu}_1 \approx \frac{1}{n} \sum_{i = 1}^{n} \bigg[ {\E}(Y_i \vert \bX_i, D_i = 1) + \frac{D_i}{\pi_{i}^{*}} \{Y_i - {\E}(Y_i \vert \bX_i, D_i = 1) \} \bigg],\label{eq:expansion-OR}\ee
where $\pi_{i}^{*}$ is the probability limit of $\widetilde{\pi}_i$. 

Alternatively, suppose one of the working PS models in (\ref{eq:MPS}), say $\pi_1(\bX_i, \bgamma_1)$, is correctly specified, but the OR model is subject to misspecification. In this case, we can replace $\widehat{\bb}(\bX_i)$ and $\widehat{\bpi}_i$ in (\ref{eq:expansion}) by their probability limits since ${\E}\{(D/\pi(\bX) - 1)u(\bX)\} = 0$ for any integrable function $u(\bX)$. This gives:
\bea
        \widehat{\mu}_1
&\approx&
\frac{1}{n} \sum_{i = 1}^{n} \bigg\{ \frac{D_i}{\widetilde{\pi}_i} - \frac{D_i}{\pi_1(\bX_i, \bgamma_{1, \ast})} \bigg\} \{Y_i - \beta_{0, \ast} - \bbeta_{1, \ast}^{\T} \bb(\bX_i) - \bbeta_{2, \ast}^{\T} \bpi_{i, \ast}\} \label{eq:expansion-PS} \\
&+&
\frac{1}{n} \sum_{i = 1}^{n} \bigg[ \frac{D_i Y_i}{\pi_1(\bX_i, \bgamma_{1, \ast})} - \frac{D_i - \pi_1(\bX_i, \bgamma_{1, \ast})}{\pi_1(\bX_i, \bgamma_{1, \ast})} \{\beta_{0, \ast}+\bbeta_{1, \ast}^{\T} \bb(\bX_i) + \bbeta_{2, \ast}^{\T} \bpi_{i, \ast} \} \bigg], \nn
\eea
where $\bpi_{i, \ast}^{\T}=(\pi_1(\bX_i, \bgamma_{1,*}),\dots,\pi_q(\bX_i, \bgamma_{q,*}))$ contains the limits of the PS models and $\bb_{\ast}(\bX_i)$ is the augmented covariate functions using probability limits of PS models for $i=1,\dots,n$. Using Taylor's expansion, the first term in (\ref{eq:expansion-PS}) can be decomposed as
\bea
&& \frac{1}{n} \sum_{i = 1}^{n} - \frac{D_i \widetilde{\ba}_i^{\T}}{\pi_1(\bX_i, \widehat{\bgamma}_{1})^2}(\widetilde{\blambda} - \blambda_{\ast}) \{Y_i - \beta_{0, \ast}- \bbeta_{1, \ast}^{\T} \bb(\bX_i) - \bbeta_{2, \ast}^{\T} \bpi_{i, \ast}\} \label{eq:outcome-bound-1} \\
&+&
\frac{1}{n} \sum_{i = 1}^{n} - \frac{D_i \pi_1^{(1)}(\bX_i, \bgamma_{1, \ast})^{\T}}{\pi_1(\bX_i, \bgamma_{1, \ast})^2} (\widehat{\bgamma}_1 - \bgamma_{1, \ast}) \{Y_i - \beta_{0, \ast}- \bbeta_{1, \ast}^{\T} \bb(\bX_i) - \bbeta_{2, \ast}^{\T} \bpi_{i, \ast}\}, \ \ \ \ \ \ \ 
\label{eq:outcome-bound-2}
\eea
where $\widetilde{\blambda} = (\widetilde{\lambda}_0, \widetilde{\blambda}_1^{\T}, \widetilde{\blambda}_2^{\T})^{\T}$, $\blambda_{\ast} = (\lambda_{0, \ast}, \blambda_{1, \ast}^{\T}, \blambda_{2, \ast}^{\T})^{\T}$, and $\widetilde{\ba}_i = (1, (\widehat{\bpi}_i - n_1 / n \mathbf{1}_q)^{\T}, \widetilde{\bb}_i^{\T})^{\T}$. The two terms in (\ref{eq:outcome-bound-1}) and (\ref{eq:outcome-bound-2}) are negligible due to the construction of the AOR with weights $\{\widetilde{\pi}_i^{-2}\}_{i=1}^{n}$ in (\ref{eq:EL-outcome}). Notably, the influence functions of $\widehat{\mu}_1$ are the same in (\ref{eq:expansion-OR}) and the second term of (\ref{eq:expansion-PS}), which establishes the validity of multiply robust confidence intervals for $\mu_1$.

From (\ref{eq:outcome-bound-1}) and (\ref{eq:outcome-bound-2}), the estimation of the nuisance parameters $\{\bgamma_{k, \ast}\}_{k = 1}^{q}$, $\lambda_{0}$, $\blambda_{1}$, and $\blambda_{2}$ does not impact the inference of the target parameter $\mu_1$, nor does the estimation of the regression coefficients $\bbeta_{1}$ and $\bbeta_{2}$ from (\ref{eq:expansion}). The proposed soft calibration and weighted AOR are constructed so that all high-dimensional nuisance parameters are Neyman orthogonal to the ATE. This is achieved by carefully designing the regularized estimation in (\ref{eq:EL-pi}) and (\ref{eq:EL-outcome}) based on KKT conditions. 

\end{document}